\documentclass[aip,jmp,preprint]{revtex4-1}

\usepackage{amsmath}
\usepackage{amssymb}
\usepackage{hyperref}
\usepackage[dvips]{graphicx}
\usepackage{color}
\begin{document}

\title{Nonrelativistic quantum dynamics on a cone with and
  without a constraining potential}

\author{C. Filgueiras}
\email{cleversonfilgueiras@yahoo.com.br}
\affiliation{Departamento de F\'{\i}sica,
  Universidade Federal de Campina Grande,
  Caixa Postal 10071,
  58109-970 Campina Grande-PB, Brazil
}

\author{E. O. Silva}
\email{edilbertoo@gmail.com}
\affiliation{
  Departamento de F\'{i}sica,
  Universidade Federal do Maranh\~{a}o,
  Campus Universit\'{a}rio do Bacanga,
  65085-580 S\~{a}o Lu\'{i}s-MA, Brazil}

\author{F. M. Andrade}
\email{fmandrade@uepg.br}
\affiliation{
  Departamento de Matem\'{a}tica e Estat\'{i}stica,
  Universidade Estadual de Ponta Grossa,
  84030-900 Ponta Grossa-PR, Brazil
}

\date{\today}

\begin{abstract}
In this paper we investigate the bound state problem of
nonrelativistic quantum particles on a conical surface.
This kind of surface appears as a  topological defect in
ordinary semiconductors as well as in graphene sheets.
Specifically, we compare and discuss the results stemming from
two different approaches.
In the first one, it is assumed that the charge carriers are
bound to the surface by a constraining potential, while the
second one is based on the Klein-Gordon type equation on
surfaces, without the constraining potential.
The main difference between both theories is the
presence/absence of a potential which contains the mean
curvature of a given surface.
This fact changes the dependence of the
bound states on the angular momentum $l$.
Moreover, there are bound states that are absent in the
Klein-Gordon theory, which instead appear in the
Schr\"{o}dinger one.
\end{abstract}

\pacs{03.65.Ge, 03.65.Db, 98.80.Cq, 11.27.+d}
\maketitle

\section{Introduction}

\label{sec:introduction}
Quantum mechanics in two dimensions is a subject of
great interest and has received much attention over the latest
years \cite{PU.2005.48.953}.
One of the main interest in this field is the study of
curvature effects on a two dimensional electron gas (2DEG).
Curvature effects on the conductance
\cite{PRB.1995.52.R8646,PB.1998.249.377,PRB.2007.75.205309,
PRB.2008.78.115326}
and curvature effects on the magnetization and persistent
currents \cite{EPJB.2003.36.183} are some phenomena studied in
2DEG's.
Usually, before considering any application, the bound state
problems of non planar 2DEG's are investigated
\cite{PRB.2002.66.205308,PRA.2006.73.012102,APB.2011.523.898}.

In 1981, R.C.T da Costa published a paper deriving the
Schr\"{o}dinger equation of a free particle constrained to move
on a curved surface \cite{PPA.1981.23.1982} (the same problem
in presence of external magnetic and electric
fields was addressed in Ref. \cite{PRL.2008.100.230403} by
Ferrari and Cuoghi).
He considered the constraint imposed by the action of an
external potential, limiting the motion of non-interacting
electrons to a thin interface with constant thickness $d$.
This way, the normal modes are separated from the ones along the
surface, in such way the Schr\"{o}dinger equation for the normal
modes(perpendicular to the interface) is
\begin{equation}
-\frac{\hbar^2}{2M}
\frac{\partial^2\chi_n}{\partial q_n}+
V(q_n)\chi_n=E_n\chi_n\;,
\label{eq:normal}
\end{equation}
where $q_n$ is the normal coordinate, and $V(q_n)$ is the
potential that confines the particle to the thin interface.
The longitudinal modes are obtained from the Schr\"{o}dinger
equation for a spinless particle written in the coordinates of a
curved surface, that is
\begin{equation}
  \frac{1}{2M}
  \left[-
    \frac{\hbar^2}{\sqrt{g}} \partial_\mu
    \left( \sqrt{g} g^{\mu\nu} \partial_{\nu}
    \right)
  \right]  \Psi+V_{\rm daCosta}\Psi=E\Psi,
  \label{eq:hamilt}
\end{equation}
where $g^{\mu\nu}$ is the contravariant component of the metric
tensor of the manifold, $g=\det g_{\mu\nu}$, and $V_{\rm daCosta}$
is a scalar geometric potential given by
\begin{equation}
V_{\rm daCosta}=-\frac{\hbar^{2}}{2M}
\left(\mathrm{M}^{2}-\mathrm{K}\right).
\label{eq:geometric}
\end{equation}
In this expression,
$\mathrm{M}=\left(\kappa_{1}+\kappa_{2}\right)/2$ is the mean
curvature, and $\mathrm{K}=\kappa_{1}\kappa_{2}$ is the Gaussian
curvature of the surface, while $\kappa_{1}$ and $\kappa_{2}$
are the principal curvatures of the surface.
The da Costa's theory is applied to a surface making
$d \rightarrow 0$.
In this theory, possible applications to a
bilayer graphene were investigated in Refs.
\cite{PRB.2009.80.153405,PRB.2009.79.033404,PRA.2010.81.014102}.
Curvature-induced  $p-n$ junctions in bent sheets were predicted
in Ref. \cite{PRB.2009.80.153405}, while an analog quantum
Hall effect due to the geometry of a helicoidal ribbon was
reported in Ref.  \cite{PRB.2009.79.033404}.

In Ref. \cite{PLA.2011.375.448}, the authors pointed out
that the intrinsic second order Dirac theory on a surface in
ordinary three-dimensional space exhibits a new scalar geometric
potential.
It is induced by the interaction between the intrinsic spin and
the surface geometry.
This term is absent in the Schr\"{o}dinger theory based on the
dimensional reduction framework of da Costa's approach.
The main difference is the absence of confinement, expressed by
Eq. \eqref{eq:normal}.
In $(2+1)$-dimensions, the squared Dirac equation is given by
\begin{equation}
\left(-D_t^2+D_{\parallel}^2+
\frac{1}{8}\mathrm{R}_{ABCD}\gamma^A \gamma^B \gamma^C
\gamma^D-M^2\right)\Psi=0\;,
\label{eq:dirac}
\end{equation}
where $\Psi$ is the Dirac spinor field, $D_{\parallel}^2$ is the
tangential surface component of the kinetic operator, the
matrices $\gamma$ obey local Clifford algebra, and
$\mathrm{R}_{ABCD}$ are the components of the Riemann
curvature.
The low energy limit of the massive Dirac theory, neglecting all
the spin-connection terms, is given by the following
Klein-Gordon (K-G) type equation
\begin{equation}
  \left(
    -\partial^2_{t}+\nabla_{\parallel}^2-
    \frac{1}{4}\mathrm{R}-M^2
  \right)\psi=0.
  \label{eq:diracexp}
\end{equation}
with $\psi$ being assumed as a definite spin-state, and
$\nabla_{\parallel}$ being the usual covariant derivative acting
on a scalar function.
We assume energy eigenstates and denote the total energy of the
particle by $E$.
Considering the first order in an $1/M$-expansion and
reinstating $\hbar$, the equation above will take the following
form,
\begin{equation}
  -\frac{\hbar^2}{2M}\nabla_{\parallel}^2\psi+
  \frac{\hbar^2}{4M}\mathrm{R}\psi=E_c \psi \;.
  \label{eq:diract}
\end{equation}
which is the Sch\"{o}redinger equation corresponding to the K-G
type equation above.
In this equation, $E_c$ is the classical energy measure and
$\mathrm{R}$ is the Ricci curvature scalar in the static
surface.
The spectrum of Eq. \eqref{eq:dirac} and \eqref{eq:diract} are
related by the relation
\begin{equation}
E_c=E-M\;.
\end{equation}
In two dimensions, $\mathrm{R}=2\mathrm{K}$, with $\mathrm{K}$
being the Gaussian curvature of a surface.
Then, we see that the Schr\"{o}dinger equation corresponding to
the K-G equation shows the effective potential,
\begin{equation}
V_{\rm geo}=\frac{\hbar^{2}}{2M}\mathrm{K}\;,
\label{eq:geometricdirac}
\end{equation}
for fermions in two dimensions.
As pointed out in reference \cite{PLA.2011.375.448},
$V_{\rm daCosta}$ and $V_{\rm geo}$ have profound differences.
While $V_{\rm daCosta}$ is always attractive, while $V_{\rm geo}$
can be either attractive or repulsive.
Notice that the difference between both geometric potentials is
that in $V_{\rm daCosta}$ there is a contribution from the mean
curvature ${\rm M}$ while no such contribution is present in
$V_{\rm geo}$.
In differential geometry, ${\rm K}$ is an {\it intrinsic}
property, since it can be written solely in terms of the metric
of a surface.
The mean curvature ${\rm M}$ depends on how a surface is immersed
in $3D$ and it is an {\it extrinsic} property.
Since in the da Costa's approach one has confinement, it would 
be expected the geometrical potential showing a piece containing
${\rm M}$.

In this work, we are interested in studying the bound state
problem of nonrelativistic quantum particles constrained to move
on a conical surface.
This problem was studied before in reference
\cite{AP.2008.323.3150}.
Here, we are going to investigate the same problem but now
considering the Sch\"{o}redinger equation which comes
from the first order $1/M$-expansion of the K-G equation.
We are interested in comparing compare the bound states for a
particle on a cone in both theories.
We choose this particular geometry since it is associated to
topological defects in ordinary semiconductors
\cite{AP.1992.216.1,JP.2000.12.7421,JPA.2003.36.863} as well as
in graphene sheets (multilayers) \cite{JETP.2001.73.562}.
We will see that, by considering $V_{\rm geo}$ instead of
$V_{\rm daCosta}$, the dependence of the bound states on the
angular momentum  $l$ is altered.

\section{The Conical Surface}

\begin{figure}[t]
  \centering
  \includegraphics[height=2.0cm]{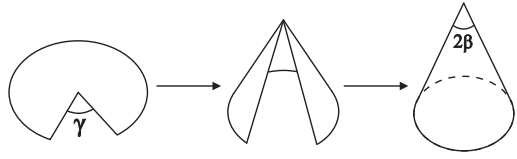}
  \caption{Conical surface of angular deficit $\gamma$.}
  \label{fig:fig1}
\end{figure}

Using polar coordinates $\rho$ and $\theta$, the following line
element
\begin{equation}
ds^{2}=d\rho^{2}+\alpha^{2}\rho^{2}d\theta^{2},
\label{eq:string1}
\end{equation}
describes a conical surface for $\rho\geq 0$ and
$0 \leq \theta< 2\pi$, describes a conical surface.
For $0<\alpha<1$ (deficit angle), the metric \eqref{eq:string1}
describes an actual cone.
Figure  \ref{fig:fig1} shows the making of a cone from a planar
sheet where an angular section was removed with posterior
identification of the edges.
If $\gamma$ is the angle that defines the removed section then
the remaining surface corresponds to an angular sector of
$2\pi\alpha=2\pi-\gamma$.
By identification of the length of the circle without the
sector, $2\pi\alpha\rho$, with the length of the complete circle
it turns out to be on the cone, $2\pi\rho\sin\beta$, we get the
relation
\begin{equation}
\alpha=\sin\beta,\label{eq:seno}
\end{equation}
where $2\beta$ is the opening angle of the cone (see
Fig. \ref{fig:fig1}.
The closer $\alpha$ gets to $1$ (or,  equivalently, $2\beta$ to
$\pi$) the flatter is the cone. For $\alpha=1$ the cone turns
into a plane.
If $\alpha>1$ (proficit angle), relation \eqref{eq:string1}
still holds and the conical surface corresponds to the insertion
of a sector (\textit{i.e.} $2\beta>\pi$).
We call the resulting surface an anti-cone.

For a cone \cite{EPL.2007.80.46002},
\begin{equation}
\mathrm{K}=\left(\frac{1-\alpha}{\alpha}\right)
\frac{\delta(\rho)}{\rho},
\label{eq:gauss1}
\end{equation}
and
\begin{equation}
\mathrm{M}=\frac{\sqrt{1-\alpha^{2}}}{2\alpha\rho}\;.
\end{equation}
Notice that the $\delta$-function singularity in the Gaussian
curvature corresponds cone tip.
In the Schr\"{o}dinger theory, a particle whose motion is
confined to the conical surface is subjected to a resultant
potential, given by
\begin{eqnarray}
V_{\rm daCosta}=-\frac{\hbar^{2}}{2M}
\left[
\left(\frac{1-\alpha^{2}}{4\alpha^2\rho^2}\right) -
\left(\frac{1-\alpha}{\alpha}\right)
\frac{\delta(\rho)}{\rho}\right].
\end{eqnarray}
There is also a contribution from a inverse squared distance
interaction.
The Schr\"{o}dinger equation for the particle, in this case,
is
\begin{equation}
  -\frac{\hbar^{2}}{2M}
  \left[\frac{1}{\rho}\frac{\partial}{\partial \rho}
    \left(\rho \frac{\partial}{\partial \rho}\right) +
    \frac{1}{\alpha^{2}\rho^{2}}
    \frac{\partial^{2}}{\partial \theta^{2}}+
    \left({\frac{1-\alpha^{2}}{4\alpha^2}}\right)\frac{1}{\rho^2}-
    \left(\frac{1-\alpha}{\alpha}\right)
    \frac{\delta(\rho)}{\rho}\right]\psi=E\psi.
\label{eq:full1}
\end{equation}
On the other hand, the K-G theory yields
\begin{equation}
  -\frac{\hbar^{2}}{2M}
  \left[\frac{1}{\rho}\frac{\partial}{\partial \rho}
    \left(\rho \frac{\partial}{\partial \rho}\right) +
    \frac{1}{\alpha^{2}\rho^{2}}
    \frac{\partial^{2}}{\partial \theta^{2}}-
    \left(\frac{1-\alpha}{\alpha}\right)
    \frac{\delta(\rho)}{\rho}
  \right]\psi=E\psi.
\label{eq:full2}
\end{equation}
Notice now that the contribution from an inverse squared
distance interaction does not appear.
This will lead to profound changes in the energy spectrum of
particles on a cone.

Now, we decompose the Hilbert space
$\mathcal{H}=L^{2}(\mathbb{R}^{2}$
with respect to the angular momentum
$\mathcal{H}=\mathcal{H}_{\rho}\otimes\mathcal{H}_{\theta}$, where
$\mathcal{H}_{\rho}=L^{2}(\mathbb{R}^{+},\rho d\rho)$ and
$\mathcal{H}_{\theta}=L^{2}(\mathcal{S}^{1},d\theta)$, with
$\mathcal{S}^{1}$ denoting the unit sphere in $\mathbb{R}^{2}$.
We have the fact that the
$-\frac{\partial^{2}}{\partial \theta^{2}}$
is essentially  self-adjoint in $L^{2}(S^{1},d\theta)$
\cite{PRA.2002.66.032118,Book.1975.Reed.II}.
So, putting the wave function in the form
\begin{equation}
\psi(\rho,\theta)=\Phi_{E}(\rho)e^{il\theta},
\end{equation}
with $l\in\mathbb{Z}$, both Eqs. \eqref{eq:full1} and
\eqref{eq:full2} can be written in a general form, that is,
\begin{equation}
  -\frac{\hbar^{2}}{2M}
    \left[
      \frac{1}{\rho}\frac{\partial}{\partial \rho}
      \left(\rho \frac{\partial}{\partial \rho}\right)-
      \frac{\mu^2}{\rho^{2}}-
    \left(\frac{1-\alpha}{\alpha}\right)
    \frac{\delta(\rho)}{\rho}
    \right]
    \Phi_{E}=E\Phi_{E}.
  \label{eq:ideal}
\end{equation}
Note that in the Schr\"{o}dinger theory it holds
\begin{equation}
  \mu^{2}=
  \frac{l^{2}}{\alpha^{2}}-
  \frac{(1-\alpha^{2})}{4\alpha^{2}},
  \label{eq:nu2}
\end{equation}
while
\begin{equation}
\mu^2=\frac{l^2}{\alpha^2}\;,  \label{eq:mudirac}
\end{equation}
in the K-G theory.
In the case of having a proficit angle, the relation
\eqref{eq:seno} should be written as
\begin{equation}
\alpha=\sinh\beta\;,
\end{equation}
since in this case we must have $\alpha>1$.
This is consistent with the fact that now the cone has a
negative curvature, that is, the surface takes a saddle-like
form.
The mean curvature is now given by
\begin{equation}
\mathrm{M}=\frac{\sqrt{\alpha^{2}-1}}{2\alpha\rho}.
\end{equation}
In this case, we have
\begin{equation}
\mu^{2}=\frac{l^{2}}{\alpha^{2}}
-\frac{\left(\alpha^{2}-1\right)}{4\alpha^{2}}\;.
\label{eq:nu3}
\end{equation}
In what follows, we replace the Gaussian curvature contribution
to the potential by a short-ranged potential supported inside a
region with radius $a$, which is small as  compared  to the
overall dimension of the system.
This way, we have $U_{\rm short}(\rho\geq a)=0$.
The Schr\"{o}dinger equation for the particle, in this case, is
\begin{equation}
  -\frac{\hbar^{2}}{2M}
  \left[\frac{1}{\rho}\frac{\partial}{\partial \rho}
    \left(\rho \frac{\partial}{\partial \rho}\right) -
    \frac{\mu^2}{\rho^{2}}
  \right]\Phi_{E}
  +\frac{\hbar^{2}}{2M}U_{\rm short}(\rho)\Phi_{E}=E\Phi_{E},
  \label{eq:full}
\end{equation}
Here, $U_{\rm short}(\rho)$ turns into the $\mathrm{K}$ of
Eq. \eqref{eq:gauss1} in the limit  $a\rightarrow 0$.

The Gaussian curvature is related to the topology of the cone,
so is $U_{\rm short}(\rho)$.
Then, it can be modeled by boundary conditions
\cite{CMP.1991.139.103}.
Moreover, from the {\it Gauss-Bonnet theorem}, we have
\cite{PLA.2005.342.237}
\begin{equation}
\int\int_{s}{\rm K}dA=\int\int_{s}U_{\rm short}dA=2\pi(1-\alpha)
\label{eq:gaussbon}\;.
\end{equation}
This fact will be considered bellow.
The problem  \eqref{eq:full} is replaced by
\begin{equation}
  \label{eq:idealized}
  -\frac{\hbar^{2}}{2M}
  \left[
    \frac{1}{\rho}\frac{\partial}{\partial \rho}
    \left(\rho \frac{\partial}{\partial \rho}\right)
    -\frac{\mu^2}{\rho^{2}}
  \right]\Phi_{\eta,E}=E \Phi_{\eta,E},
\end{equation}
with $\Phi_{\eta,E}$ labeled by a parameter $\eta$ which is
related to the behavior of the wave function in the limit
$\rho\rightarrow a$.
Next, we have to discover which  boundary conditions are allowed
to Eq. \eqref{eq:ideal}.
This is performed via the the self-adjoint extension approach
\cite{Book.1975.Reed.II,AJP.2001.69.322}.

Before proceeding to find the solutions of Eq. \eqref{eq:ideal},
we will revise the bound states addressed in the Schr\"{o}dinger
theory.
The details can be found in reference \cite{AP.2008.323.3150}.

\section{Bound States - da Costa approach}
\label{sec:bsschor}

We start by revising the bound states in the Schr\"{o}dinger
theory.
For Eq. \eqref{eq:ideal}, two possible solutions rise up.
The first one, corresponding to $\mu^2\leq0$, is
\begin{equation}
\Phi_{E}(\rho)=K_{i|\mu|}\left(\frac{\rho}{\hbar}\sqrt{-2mE}\right),
\end{equation}
where $K_{\nu}(\rho)$ is the {\it Bessel function of third kind},
$\left|\mu\right|<1$, and the energies given by
\begin{equation}
  E=-\frac{2\hbar^{2}}{Ma^{2}}
  \exp\left[
    \frac{2}{\mu}\cot^{-1}
    \left(\frac{1-\alpha}{\alpha\mu}-
      \frac{\mu}{2}
    \right)-2\gamma_{e}
  \right],
  \label{eq:novel}
\end{equation}
here $\gamma_{e}$ is the Euler-Mascheroni constant.

These results corresponds to $\alpha<1$.
In this case, the mean curvature contributes attractively, while
the Gaussian curvature yields a repulsive short-ranged
potential.
Eq. \eqref{eq:nu2} implies that the only allowable value for the
angular momentum is $l=0$, meaning that we have a single bound
state.
For $\alpha>1$, the relation \eqref{eq:nu3} holds.
The mean curvature contributes attractively again and the
Gaussian curvature now leads to an attractive short-ranged
potential.
We have a single bound state for $l =0$, as before.
On the other hand, for $\mu^2\geq0$, the wave solutions are
\begin{equation}
\Phi_{E}(\rho)=K_{\mu}\left(\frac{\rho}{\hbar}\sqrt{-2mE}\right),
\end{equation}
for $\left|\mu\right|<1$ and, the energies are given by
\cite{note-nrd}
\begin{equation}
  \label{eq:e1}
  E= -\frac{2\hbar^{2}}{M a^{2}}
  \left[
    \frac{\Gamma(1+\mu)}{\Gamma (1-\mu)}
    \left(
      \frac
      {\mu^{2}+2\left(\frac{1-\alpha}{\alpha}\right)-2\mu}
      {\mu^{2}+2\left(\frac{1-\alpha}{\alpha}\right)+2\mu}
    \right)
  \right]^{1/\mu}.
\end{equation}
We have bound states for
\begin{equation}
\begin{array}{rrrrl}
1          & <\alpha & \leq \sqrt{5}\;\;
& \text{at} & l=\pm 1, \\
\sqrt{17/5}& <\alpha & \leq \sqrt{17}
& \text{at} & l=\pm 1 \text{ and }\pm 2,\\
\sqrt{37/5}& <\alpha & <    \sqrt{37}
& \text{at} & l=\pm 1,\pm2, \text{ and } \pm3,
\end{array}
\end{equation}
and so on.

In the next section, we solve the same problem in the K-G theory
context, comparing the results with the ones discussed here.

\section{Bound states - Schr\"{o}dinger equation
from the K-G theory}
\label{sec:bsdirac}

We now look for the bound states solving the Schr\"{o}dinger
equation which comes from the K-G theory.
As discussed above, we can solve this problem using the
self-adjoint extension approach.
From  Eq. \eqref{eq:idealized}, we have
\begin{equation}
\label{eq:hfull}
\mathcal{H} \Phi_{\eta,E}=\kappa^{2} \Phi_{\eta,E},
\end{equation}
where
\begin{equation}
  \mathcal{H}=-\frac{\hbar^{2}}{2M}
  \left[\frac{1}{\rho}
    \frac{\partial}{\partial \rho}
    \left(\rho \frac{\partial}{\partial \rho}\right)-
    \frac{\mu^2}{\rho^{2}}
  \right].
\end{equation}
Eq. \eqref{eq:hfull} is the modified Bessel equation.
In it, $\kappa^2=-2ME/\hbar^2>0$, since we are looking for bound
states.

Notice that, different from the Schr\"{o}dinger theory, the
inverse squared potential here is always attractive.
This happens because, from Eq. \eqref{eq:mudirac}, the only
possible case is $\mu^2\geq0$.
For  $\alpha<1$ (deficit angle), the Gaussian curvature is
positive and the short ranged potential is repulsive, meaning
that we have only scattering states.
On the other hand, when  $\alpha>1$ the Gaussian curvature is
negative and such short ranged potential is negative, implying
attractiveness.
This is, therefore, the only case where bound states can exist.

This way, we can see the first difference between the two
theories as follows. In the Schr\"{o}dinger one, we can have
bound states for either $\mu^2>0$ or $\mu^2<0$
(there is no possibility of having $\mu=0$), regardless if we
have a deficit ($\alpha<1$) or a proficit ($\alpha>1$) angle.
For the K-G theory the only possibility is for a proficit angle,
since we have only the case $\mu^2\geq0$. Note that this
difference relies in the fact that the contribution from mean
curvature is now absent in this theory.

We now proceed in order to find the bound states in the K-G
theory. 
First of all, we must determine the full domain of $\mathcal{H}$
in $L^{2}(\mathcal{R}^{+},\rho  d\rho)$.
In doing so, we have to find its deficient subspaces.
This is done solving the eigenvalue equation
\begin{equation}
  \label{eq:eigen}
  \mathcal{H}^{\dagger}\Phi_{\pm}=\pm ik_{0}\Phi_{\pm},
\end{equation}
where $\mathcal{H}^{\dagger}=\mathcal{H}$.
The general solution for this equation is given in terms of
\textit{modified Bessel functions}, denoted by $I_\mu$ and
$K_\mu$. However, only $K_\mu$ is square integrable in all
space. 
Then, we have as solutions of Eq. \eqref{eq:eigen} the wave
functions
\begin{equation}
  \Phi_{\pm}(\rho)=\text{const.}\;
  K_{\mu}\left(\frac{\rho}{\hbar}\sqrt{\mp\varepsilon}
\right)\;,  \label{eq:subspace}
\end{equation}
with $\mu^2<1$ and $\varepsilon=2iMk_{0}$.
The dimension of such deficient subspaces is
$(n_{+},n_{-})=(1,1)$.
Then, the wave functions $\Phi_{\eta,E}$ are written as
\begin{equation}
\Phi_{\eta,E}(\rho)=\chi_{\mu}(\rho)+
C\left[
  K_{\mu}\left(\frac{\rho}{\hbar}\sqrt{-\varepsilon}\right)+
  e^{i\eta}K_{\mu}\left(\frac{\rho}{\hbar}\sqrt{\varepsilon}\right)
\right],
\label{eq:domain}
\end{equation}
where $\chi_{\mu}(\rho)$, with $\chi_{\mu}(a)=\dot{\chi}_{\mu}(a)=0$.
The last term in Eq. \eqref{eq:domain} gives the correct
behavior of the wave function when $\rho=a$.
The parameters $\eta\;(mod2\pi)$ and $k_0$ represent the
\textit{a priori} choices of boundary conditions.
The $U_{\rm short}$ potential determines these parameters without
ambiguity. 
This is done by finding a fitting formula for $\eta$
\cite{CMP.1991.139.103}: we write $E=0$ or the static solution
for the problem, that is 
\begin{equation}
  -\frac{\hbar^{2}}{2M}
  \left[\frac{1}{\rho}\frac{\partial}{\partial \rho}
    \left(\rho \frac{\partial}{\partial \rho}\right) -
    \frac{\mu^2}{\rho^{2}}
  \right]\Phi_{0}
  +\frac{\hbar^{2}}{2M}U_{\rm short}(\rho)\Phi_{0}=0.
  \label{eq:statictrue}
\end{equation}

Now, we require the continuity for the logarithmic derivative
\begin{equation}
\frac{\rho}{\Phi_{0}}\frac{d \Phi_{0}}{d\rho}\Big|_{\rho=a}=
\frac{\rho}{\Phi_{\eta,0}}\frac{d \Phi_{\eta,0}}{d\rho}\Big|_{\rho=a},
\label{eq:logder}
\end{equation}
where $\Phi_{\eta,0}$ comes from Eq. \eqref{eq:domain} for $E=0$.
The left-hand side of this equation can be achieved integrating
\eqref{eq:statictrue} from $0$ to $a$,
\begin{equation}
  \label{eq:inth0}
  -\int_{0}^{a}\frac{1}{\rho}\frac{d}{d\rho}
  \left( \rho\frac{d\Phi_{0}}{d\rho}\right)
  \rho d\rho+
    \int_{0}^{a}\Phi_{0}U_{\rm short}(\rho)\rho d\rho
  +\mu^{2}\int_{0}^{a}\frac{\Phi_{0}}{\rho^{2}}\rho d\rho=0.
\end{equation}
Now, considering that $\Phi_{0}/\rho^{2}$ does not change
significantly in the range $[0,a]$, we have
\begin{equation}
\int_{0}^{a}\frac{\Phi_{0}(\rho)}{r^{2}}\rho d\rho\approx
\frac{\Phi_{0}(a)}{a^{2}}\int_{0}^{a}\rho d\rho=\frac{\Phi_{0}(a)}{2}.
\end{equation}
From Eq. \eqref{eq:gaussbon}, we have
\begin{equation}
  \int_{0}^{a}\Phi_{0}U_{\rm short}(\rho)\rho d\rho\approx
  \Phi_{0}\int_{0}^{a}U_{\rm short}(\rho)\rho d\rho=
  \Phi_{0}\int_{0}^{a}{\rm K}\rho d\rho=
  \frac{\left(1-\alpha\right)}{\alpha}\Phi_{0}\;,
  \label{eq:}
\end{equation}
so that
\begin{equation}
\frac{a}{\Phi_{0}(a)}\frac{d\Phi_{0}(\rho)}{dr}\Big|_{\rho=a}=
\frac{\left(1-\alpha\right)}{\alpha} +
\frac{\mu^{2}}{2}.  \label{eq:nrs}
\end{equation}

Since $a\approx 0$, the right-hand side of Eq. \eqref{eq:logder}
is calculated using the asymptotic representation for
$K_{\mu}(x)$ in the limit  $x\rightarrow 0$, given by
\begin{equation}
  K_{\nu}(x)\sim \frac{\pi}{2\sin (\pi \nu)}
  \left[
    \frac{x^{-\nu}}{2^{-\nu}\Gamma (1-\nu)}-
    \frac{x^{ \nu}}{2^{ \nu}\Gamma (1+\nu)}
  \right].
\label{eq:besselasympt}
\end{equation}
Thus, taking into account \eqref{eq:domain}, we arrive at
\begin{equation}
  \frac{a}{\Phi_{\eta,0}(a)}
  \frac{d \Phi_{\eta,0}(\rho)}{d \rho}\Big|_{\rho=a}=
  \frac{1}{\Omega_{\eta}(a)}
  \frac{d \Omega_{\eta}(\rho)}{d \rho}\Big|_{\rho=a},
  \label{eq:dright}
\end{equation}
where
\begin{equation}
  \label{eq:omegas}
  \Omega_{\eta}(\rho)=
  \left[
    \frac
    {\left(\rho\sqrt{-\varepsilon}\right)^{-\mu}}
    {2^{-\mu}\Gamma (1-\mu)}-
    \frac
    {\left(\rho\sqrt{-\varepsilon}\right)^{\mu}}
    {2^{\mu}\Gamma (1+\mu)}
  \right]+
  e^{i\eta}
  \left[
    \frac
    {\left(\rho\sqrt{\varepsilon}\right)^{-\mu}}
    {2^{-\mu}\Gamma (1-\mu)} -
    \frac
    {\left(\rho\sqrt{\varepsilon}\right)^{\mu}}
    {2^{\mu}\Gamma (1+\mu)}\right]
\end{equation}
By inserting Eqs. \eqref{eq:nrs} and \eqref{eq:dright} into
Eq. \eqref{eq:logder}, we find
\begin{equation}
  \label{eq:saepapprox}
  \frac{1}{\Omega_{\eta}(a)}
  \frac{d \Omega_{\eta}(\rho)}{d \rho}\Big|_{\rho=a}=
  \frac{\left(1-\alpha\right)}{\alpha}+\frac{\mu^{2}}{2},
\end{equation}
which gives us the parameter $\eta$ in terms of the physics of
the problem, that is, the correct behavior of the wave functions
when $\rho\rightarrow a$, or the coupling between the
short-ranged potential $U_{\text{short}}(\rho)$ and the wave
functions. 

Next, we will find the bound states of the Hamiltonian
$\mathcal{H}$ and using Eq. \eqref{eq:saepapprox}, the spectrum
will be determined without any arbitrary parameter
\cite{PRD.2012.85.041701,AP.2010.325.2529}. 
In doing so, we can go back to
Eq. \eqref{eq:idealized}, whose general solution is
\begin{equation}
  \label{eq:sver}
  \Phi_{\eta,E}(\rho)=
  K_{\mu}\left(\frac{\rho}{\hbar}\sqrt{-2ME}\right).
\end{equation}
Since these solutions belong to the domain of $\mathcal{H}$, it
has the form of Eq. \eqref{eq:domain}, that is,
\begin{equation}
  \label{eq:domainE}
  \Phi_{\eta,E}(\rho)=\chi_{\mu}(\rho)+
  C\left[
    K_{\mu}(\frac{\rho}{\hbar}\sqrt{-\varepsilon})+
    e^{i\eta}K_{\mu}(\frac{\rho}{\hbar}\sqrt{\varepsilon})
  \right],
\end{equation}
for some $\eta$ selected from the physics of the problem. 
So, we substitute \eqref{eq:sver} in Eq. \eqref{eq:domain} and
compute $a/\Phi_{\eta,E}(a) (d\Phi_{\eta,E}(\rho)/d\rho)|_{\rho=a}$,
using \eqref{eq:besselasympt}, achieving
\begin{equation}
  \label{eq:derfe}
  \frac{a}{\Phi_{\eta,E}(a)}
  \frac{d\Phi_{\eta,E}(a)}{d\rho}\Big|_{\rho=a}=
  \frac
  {\mu \left[a^{2\mu} \Gamma(1-\mu)(-M E/\hbar^{2})^{\mu}
      +2^{\mu} \Gamma(1+\mu)\right]}
  {a^{2 \mu}\Gamma(1-\mu)(-M E/\hbar^{2})^{\mu}
    -2^{\mu} \Gamma(1+\mu)}
  =\frac{1}{\Omega_{\eta}(a)}\frac{d \Omega_{\eta}(a)}{d \rho}.
\end{equation}
By using Eq. \eqref{eq:saepapprox} and solving the equation
above for $E$, we find the energy spectrum
\begin{equation}  
\label{eq:energy_KS}
  E= -\frac{2\hbar^{2}}{Ma^{2}}
  \left[
    \frac{\Gamma(1+\mu)}{\Gamma (1-\mu)}
    \left(
      \frac
      {\mu^{2}+2\left(\frac{1-\alpha}{\alpha}\right)-2\mu}
      {\mu^{2}+2\left(\frac{1-\alpha}{\alpha}\right)+2\mu}
    \right)
  \right]^{1/\mu},
\end{equation}
which is similar to that described in formula \eqref{eq:e1}.
The main difference is that now the effective angular momentum
is just $\mu=\pm\frac{l }{\alpha}$ instead of the one given by
Eqs. \eqref{eq:nu2} and \eqref{eq:nu3}.
As we noted, when $\alpha<1$, we have $U_{\rm short}>0$,
that is, there is no bound states in this case. 
On the other hand, if $\alpha>1$, $U_{\rm short}<0$ and bound
states can show up.
Because of the condition $\left|\mu\right|<1$, the value of the
proficit angle will determine which values of angular momentum
are allowed.
For example, if $\alpha=1.5$, then we must have $l =-1,0,1$. If
$\alpha=2.1$, then $l$ admits the values $-2,-1,0,1,2$, and so
on. 
Now, we summarize the results of this section, that is, 
\begin{equation}
\begin{array}{rrrrl}
1 & <\alpha & \leq 2 & \text{at} & l=0, \pm 1, \\
2 & <\alpha & \leq 3 & \text{at} & l=0, \pm 1 \text{ and }\pm 2,\\
3 & <\alpha & \leq 4 & \text{at} & l=0, \pm 1,\pm 2, \text{ and } \pm 3,
\end{array}
\end{equation}
and so on.

Notice that the dependence of the bound states on the angular
momentum $l$ changes significantly when compared to the
dependence observed in the Schr\"{o}dinger theory.
For example, the null angular momentum, $l =0$, now can exist.

Concluding this section, when $\mu^2>0$ and $\alpha>1$, the mean
curvature potential is repulsive and the short-ranged potential,
due to the Gaussian curvature, is attractive in the
Schr\"{o}dinger theory.
On the other hand, in the K-G theory, the absence of the mean
curvature potential affects drastically the bound states of a
quantum particle on a cone.
In both cases, the Gaussian curvature is the sole responsible
for the bound states.

\section{Concluding remarks}

In a previous work \cite{AP.2008.323.3150}, the quantum dynamics
of a particle on a cone was addressed in the context of the
Schr\"{o}dinger theory considering the da Costa's approach.
In this work we investigate the bound states of a quantum
particle on the same surface within the K-G theory framework.
In both theories, the conical geometry provides a
$\delta$-function interaction which can be either attractive or
repulsive, depending on the cone parameter $\alpha$.
However, in the first theory, the conical geometry introduces an
inverse squared distance potential due to the mean curvature
which leads to an effective potential that can be either
attractive or repulsive, depending on $\alpha$.
This mean curvature term does not show up when we study such
system using the K-G theory.
We saw in the previous sections that this difference implies
profound changes for the dependence of the bound states on the 
angular momentum $l$.
These results are summarized in Tables \ref{tab:table1} and
\ref{tab:table2}.

\begin{table}[t]
  \centering
  \begin{ruledtabular}
    \begin{tabular}{ccc}
      & $\alpha > 1$ & $\alpha <1$ \\
      \hline
      $\mu^{2}<0$ & $1$ bound state for $l=0$
      & $1$ bound state for $l=0$
      \\
      $\mu^{2}>0$ & bound states & scattering states for $l\neq0$
    \end{tabular}
  \end{ruledtabular}
  \caption{Summary of the results - Schr\"{o}dinger Theory}
  \label{tab:table1}
\end{table}

\begin{table}[t]
  \centering
  \begin{ruledtabular}
    \begin{tabular}{ccc}
      & $\alpha > 1$ & $\alpha <1$ \\
      \hline
      $\mu^{2}\geq0$ & bound states & scattering states
    \end{tabular}
  \end{ruledtabular}
  \caption{Summary of the results - K-G Theory}
  \label{tab:table2}
\end{table}

From Table \ref{tab:table1}, we see that, for $\alpha>1$, the
attractive short-ranged potential (Gaussian curvature)
guarantees the existence of bound states.
For $\alpha<1$ (repulsive short-ranged potential), an
attractive effective potential yields a bound state for $l=0$.
However, for $l\neq 0$, we have only repulsive potentials and no
bound states appear.

In Table \ref{tab:table2}, the absence of the mean curvature
makes the inverse square potential always repulsive, so the
Gaussian curvature is the unique responsible for the existence
of bound states.
They exist only when we have a proficit angle $\alpha$, that is,
$\alpha>1$.
No bound states appear in the actual cone ($\alpha<1$).
This is in great contrast with the results which comes from the
Schr\"{o}dinger theory.

In order to investigate the behavior of fermions in two
dimensions, it is important to explore phenomena like the
quantum Hall effect, magnetization and persistent currents in
two dimensional semiconductors.
For a cone, these two last physical quantities were investigated
in reference \cite{EPL.2007.79.57001}.
The authors did not take into account either $V_{\rm daCosta}$
or $V_{\rm geo}$.
But their results are correct within the K-G theory, since no
potential containing the mean curvature of a cone appears and
the electrons are localized on a ring far from the cone
apex.
However, if we consider the Schr\"{o}dinger theory with the da
Costa's approach, the effective angular momentum now depends on an
extra piece which comes from $V_{\rm daCosta}$, as we saw above.
So, measurements on the persistent current on conical surfaces
can help to decide which theory is suitable to describe fermions
in two dimensions.
Perhaps, both theories can be used depending on which material
we have in hands.
We must mention do not not apply to a monolayer graphene, where
fermions are in fact described by a massless Dirac theory.
Although, they could be important for quantum systems in two
dimensional surfaces consisting of more than one layer, as a
bilayer graphene sheet.

As a final word, it is clear that both theories provide,
theoretically, contradictory physical results.
We have discussed this difference here, for quantum particles on
a conical surface.
The K-G theory can also change the scenario when we consider
possible applications to a bilayer graphene.

\section*{Acknowledgments}

The authors would like to thank M. M. Ferreira Jr. for critical
reading the manuscript.
E. O. Silva acknowledges research grants by CNPq-(Universal/2011)
and C. Filgueiras by CAPES (Nanobiotec)/CNPq(Universal/2011).


\begin{thebibliography}{30}%
\makeatletter
\providecommand \@ifxundefined [1]{%
 \@ifx{#1\undefined}
}%
\providecommand \@ifnum [1]{%
 \ifnum #1\expandafter \@firstoftwo
 \else \expandafter \@secondoftwo
 \fi
}%
\providecommand \@ifx [1]{%
 \ifx #1\expandafter \@firstoftwo
 \else \expandafter \@secondoftwo
 \fi
}%
\providecommand \natexlab [1]{#1}%
\providecommand \enquote  [1]{``#1''}%
\providecommand \bibnamefont  [1]{#1}%
\providecommand \bibfnamefont [1]{#1}%
\providecommand \citenamefont [1]{#1}%
\providecommand \href@noop [0]{\@secondoftwo}%
\providecommand \href [0]{\begingroup \@sanitize@url \@href}%
\providecommand \@href[1]{\@@startlink{#1}\@@href}%
\providecommand \@@href[1]{\endgroup#1\@@endlink}%
\providecommand \@sanitize@url [0]{\catcode `\\12\catcode `\$12\catcode
  `\&12\catcode `\#12\catcode `\^12\catcode `\_12\catcode `\%12\relax}%
\providecommand \@@startlink[1]{}%
\providecommand \@@endlink[0]{}%
\providecommand \url  [0]{\begingroup\@sanitize@url \@url }%
\providecommand \@url [1]{\endgroup\@href {#1}{\urlprefix }}%
\providecommand \urlprefix  [0]{URL }%
\providecommand \Eprint [0]{\href }%
\providecommand \doibase [0]{http://dx.doi.org/}%
\providecommand \selectlanguage [0]{\@gobble}%
\providecommand \bibinfo  [0]{\@secondoftwo}%
\providecommand \bibfield  [0]{\@secondoftwo}%
\providecommand \translation [1]{[#1]}%
\providecommand \BibitemOpen [0]{}%
\providecommand \bibitemStop [0]{}%
\providecommand \bibitemNoStop [0]{.\EOS\space}%
\providecommand \EOS [0]{\spacefactor3000\relax}%
\providecommand \BibitemShut  [1]{\csname bibitem#1\endcsname}%
\let\auto@bib@innerbib\@empty
\bibitem [{\citenamefont {Magarill}\ \emph {et~al.}(2005)\citenamefont
  {Magarill}, \citenamefont {Chaplik},\ and\ \citenamefont
  {Entin}}]{PU.2005.48.953}%
  \BibitemOpen
  \bibfield  {author} {\bibinfo {author} {\bibfnamefont {L.~I.}\ \bibnamefont
  {Magarill}}, \bibinfo {author} {\bibfnamefont {A.~V.}\ \bibnamefont
  {Chaplik}}, \ and\ \bibinfo {author} {\bibfnamefont {M.~V.}\ \bibnamefont
  {Entin}},\ }\href {\doibase 10.1070/PU2005v048n09ABEH005730} {\bibfield
  {journal} {\bibinfo  {journal} {Phys. Usp.}\ }\textbf {\bibinfo {volume}
  {48}},\ \bibinfo {pages} {953} (\bibinfo {year} {2005})}\BibitemShut
  {NoStop}%
\bibitem [{\citenamefont {Foden}\ \emph {et~al.}(1995)\citenamefont {Foden},
  \citenamefont {Leadbeater},\ and\ \citenamefont
  {Pepper}}]{PRB.1995.52.R8646}%
  \BibitemOpen
  \bibfield  {author} {\bibinfo {author} {\bibfnamefont {C.~L.}\ \bibnamefont
  {Foden}}, \bibinfo {author} {\bibfnamefont {M.~L.}\ \bibnamefont
  {Leadbeater}}, \ and\ \bibinfo {author} {\bibfnamefont {M.}~\bibnamefont
  {Pepper}},\ }\href {\doibase 10.1103/PhysRevB.52.R8646} {\bibfield  {journal}
  {\bibinfo  {journal} {Phys. Rev. B}\ }\textbf {\bibinfo {volume} {52}},\
  \bibinfo {pages} {R8646} (\bibinfo {year} {1995})}\BibitemShut {NoStop}%
\bibitem [{\citenamefont {Chaplik}\ \emph {et~al.}(1998)\citenamefont
  {Chaplik}, \citenamefont {Magarill},\ and\ \citenamefont
  {Romanov}}]{PB.1998.249.377}%
  \BibitemOpen
  \bibfield  {author} {\bibinfo {author} {\bibfnamefont {A.}~\bibnamefont
  {Chaplik}}, \bibinfo {author} {\bibfnamefont {L.}~\bibnamefont {Magarill}}, \
  and\ \bibinfo {author} {\bibfnamefont {D.}~\bibnamefont {Romanov}},\ }\href
  {\doibase 10.1016/S0921-4526(98)00135-5} {\bibfield  {journal} {\bibinfo
  {journal} {Physica B}\ }\textbf {\bibinfo {volume} {249–251}},\ \bibinfo
  {pages} {377 } (\bibinfo {year} {1998})}\BibitemShut {NoStop}%
\bibitem [{\citenamefont {Vorob'ev}\ \emph {et~al.}(2007)\citenamefont
  {Vorob'ev}, \citenamefont {Friedland}, \citenamefont {Kostial}, \citenamefont
  {Hey}, \citenamefont {Jahn}, \citenamefont {Wiebicke}, \citenamefont
  {Yukecheva},\ and\ \citenamefont {Prinz}}]{PRB.2007.75.205309}%
  \BibitemOpen
  \bibfield  {author} {\bibinfo {author} {\bibfnamefont {A.~B.}\ \bibnamefont
  {Vorob'ev}}, \bibinfo {author} {\bibfnamefont {K.-J.}\ \bibnamefont
  {Friedland}}, \bibinfo {author} {\bibfnamefont {H.}~\bibnamefont {Kostial}},
  \bibinfo {author} {\bibfnamefont {R.}~\bibnamefont {Hey}}, \bibinfo {author}
  {\bibfnamefont {U.}~\bibnamefont {Jahn}}, \bibinfo {author} {\bibfnamefont
  {E.}~\bibnamefont {Wiebicke}}, \bibinfo {author} {\bibfnamefont {J.~S.}\
  \bibnamefont {Yukecheva}}, \ and\ \bibinfo {author} {\bibfnamefont {V.~Y.}\
  \bibnamefont {Prinz}},\ }\href {\doibase 10.1103/PhysRevB.75.205309}
  {\bibfield  {journal} {\bibinfo  {journal} {Phys. Rev. B}\ }\textbf {\bibinfo
  {volume} {75}},\ \bibinfo {pages} {205309} (\bibinfo {year}
  {2007})}\BibitemShut {NoStop}%
\bibitem [{\citenamefont {Ferrari}\ \emph {et~al.}(2008)\citenamefont
  {Ferrari}, \citenamefont {Bertoni}, \citenamefont {Goldoni},\ and\
  \citenamefont {Molinari}}]{PRB.2008.78.115326}%
  \BibitemOpen
  \bibfield  {author} {\bibinfo {author} {\bibfnamefont {G.}~\bibnamefont
  {Ferrari}}, \bibinfo {author} {\bibfnamefont {A.}~\bibnamefont {Bertoni}},
  \bibinfo {author} {\bibfnamefont {G.}~\bibnamefont {Goldoni}}, \ and\
  \bibinfo {author} {\bibfnamefont {E.}~\bibnamefont {Molinari}},\ }\href
  {\doibase 10.1103/PhysRevB.78.115326} {\bibfield  {journal} {\bibinfo
  {journal} {Phys. Rev. B}\ }\textbf {\bibinfo {volume} {78}},\ \bibinfo
  {pages} {115326} (\bibinfo {year} {2008})}\BibitemShut {NoStop}%
\bibitem [{\citenamefont {Bulaev}\ and\ \citenamefont
  {Margulis}(2003)}]{EPJB.2003.36.183}%
  \BibitemOpen
  \bibfield  {author} {\bibinfo {author} {\bibfnamefont {D.}~\bibnamefont
  {Bulaev}}\ and\ \bibinfo {author} {\bibfnamefont {V.}~\bibnamefont
  {Margulis}},\ }\href {\doibase 10.1140/epjb/e2003-00333-x} {\bibfield
  {journal} {\bibinfo  {journal} {Eur. Phys. J. B}\ }\textbf {\bibinfo {volume}
  {36}},\ \bibinfo {pages} {183} (\bibinfo {year} {2003})}\BibitemShut
  {NoStop}%
\bibitem [{\citenamefont {Entin}\ and\ \citenamefont
  {Magarill}(2002)}]{PRB.2002.66.205308}%
  \BibitemOpen
  \bibfield  {author} {\bibinfo {author} {\bibfnamefont {M.~V.}\ \bibnamefont
  {Entin}}\ and\ \bibinfo {author} {\bibfnamefont {L.~I.}\ \bibnamefont
  {Magarill}},\ }\href {\doibase 10.1103/PhysRevB.66.205308} {\bibfield
  {journal} {\bibinfo  {journal} {Phys. Rev. B}\ }\textbf {\bibinfo {volume}
  {66}},\ \bibinfo {pages} {205308} (\bibinfo {year} {2002})}\BibitemShut
  {NoStop}%
\bibitem [{\citenamefont {Encinosa}(2006)}]{PRA.2006.73.012102}%
  \BibitemOpen
  \bibfield  {author} {\bibinfo {author} {\bibfnamefont {M.}~\bibnamefont
  {Encinosa}},\ }\href {\doibase 10.1103/PhysRevA.73.012102} {\bibfield
  {journal} {\bibinfo  {journal} {Phys. Rev. A}\ }\textbf {\bibinfo {volume}
  {73}},\ \bibinfo {pages} {012102} (\bibinfo {year} {2006})}\BibitemShut
  {NoStop}%
\bibitem [{\citenamefont {Filgueiras}\ and\ \citenamefont
  {de~Oliveira}(2011)}]{APB.2011.523.898}%
  \BibitemOpen
  \bibfield  {author} {\bibinfo {author} {\bibfnamefont {C.}~\bibnamefont
  {Filgueiras}}\ and\ \bibinfo {author} {\bibfnamefont {B.}~\bibnamefont
  {de~Oliveira}},\ }\href {\doibase 10.1002/andp.201000158} {\bibfield
  {journal} {\bibinfo  {journal} {Ann. Phys. (Berlin)}\ }\textbf {\bibinfo
  {volume} {523}},\ \bibinfo {pages} {898} (\bibinfo {year}
  {2011})}\BibitemShut {NoStop}%
\bibitem [{\citenamefont {da~Costa}(1981)}]{PPA.1981.23.1982}%
  \BibitemOpen
  \bibfield  {author} {\bibinfo {author} {\bibfnamefont {R.~C.~T.}\
  \bibnamefont {da~Costa}},\ }\href {\doibase 10.1103/PhysRevA.23.1982}
  {\bibfield  {journal} {\bibinfo  {journal} {Phys. Rev. A}\ }\textbf {\bibinfo
  {volume} {23}},\ \bibinfo {pages} {1982} (\bibinfo {year}
  {1981})}\BibitemShut {NoStop}%
\bibitem [{\citenamefont {Ferrari}\ and\ \citenamefont
  {Cuoghi}(2008)}]{PRL.2008.100.230403}%
  \BibitemOpen
  \bibfield  {author} {\bibinfo {author} {\bibfnamefont {G.}~\bibnamefont
  {Ferrari}}\ and\ \bibinfo {author} {\bibfnamefont {G.}~\bibnamefont
  {Cuoghi}},\ }\href {\doibase 10.1103/PhysRevLett.100.230403} {\bibfield
  {journal} {\bibinfo  {journal} {Phys. Rev. Lett.}\ }\textbf {\bibinfo
  {volume} {100}},\ \bibinfo {pages} {230403} (\bibinfo {year}
  {2008})}\BibitemShut {NoStop}%
\bibitem [{\citenamefont {Joglekar}\ and\ \citenamefont
  {Saxena}(2009)}]{PRB.2009.80.153405}%
  \BibitemOpen
  \bibfield  {author} {\bibinfo {author} {\bibfnamefont {Y.~N.}\ \bibnamefont
  {Joglekar}}\ and\ \bibinfo {author} {\bibfnamefont {A.}~\bibnamefont
  {Saxena}},\ }\href {\doibase 10.1103/PhysRevB.80.153405} {\bibfield
  {journal} {\bibinfo  {journal} {Phys. Rev. B}\ }\textbf {\bibinfo {volume}
  {80}},\ \bibinfo {pages} {153405} (\bibinfo {year} {2009})}\BibitemShut
  {NoStop}%
\bibitem [{\citenamefont {Atanasov}\ \emph {et~al.}(2009)\citenamefont
  {Atanasov}, \citenamefont {Dandoloff},\ and\ \citenamefont
  {Saxena}}]{PRB.2009.79.033404}%
  \BibitemOpen
  \bibfield  {author} {\bibinfo {author} {\bibfnamefont {V.}~\bibnamefont
  {Atanasov}}, \bibinfo {author} {\bibfnamefont {R.}~\bibnamefont {Dandoloff}},
  \ and\ \bibinfo {author} {\bibfnamefont {A.}~\bibnamefont {Saxena}},\ }\href
  {\doibase 10.1103/PhysRevB.79.033404} {\bibfield  {journal} {\bibinfo
  {journal} {Phys. Rev. B}\ }\textbf {\bibinfo {volume} {79}},\ \bibinfo
  {pages} {033404} (\bibinfo {year} {2009})}\BibitemShut {NoStop}%
\bibitem [{\citenamefont {Dandoloff}\ \emph {et~al.}(2010)\citenamefont
  {Dandoloff}, \citenamefont {Saxena},\ and\ \citenamefont
  {Jensen}}]{PRA.2010.81.014102}%
  \BibitemOpen
  \bibfield  {author} {\bibinfo {author} {\bibfnamefont {R.}~\bibnamefont
  {Dandoloff}}, \bibinfo {author} {\bibfnamefont {A.}~\bibnamefont {Saxena}}, \
  and\ \bibinfo {author} {\bibfnamefont {B.}~\bibnamefont {Jensen}},\ }\href
  {\doibase 10.1103/PhysRevA.81.014102} {\bibfield  {journal} {\bibinfo
  {journal} {Phys. Rev. A}\ }\textbf {\bibinfo {volume} {81}},\ \bibinfo
  {pages} {014102} (\bibinfo {year} {2010})}\BibitemShut {NoStop}%
\bibitem [{\citenamefont {Jensen}\ and\ \citenamefont
  {Dandoloff}(2011)}]{PLA.2011.375.448}%
  \BibitemOpen
  \bibfield  {author} {\bibinfo {author} {\bibfnamefont {B.}~\bibnamefont
  {Jensen}}\ and\ \bibinfo {author} {\bibfnamefont {R.}~\bibnamefont
  {Dandoloff}},\ }\href {\doibase 10.1016/j.physleta.2010.12.018} {\bibfield
  {journal} {\bibinfo  {journal} {Phys. Lett. A}\ }\textbf {\bibinfo {volume}
  {375}},\ \bibinfo {pages} {448 } (\bibinfo {year} {2011})}\BibitemShut
  {NoStop}%
\bibitem [{\citenamefont {Filgueiras}\ and\ \citenamefont
  {Moraes}(2008)}]{AP.2008.323.3150}%
  \BibitemOpen
  \bibfield  {author} {\bibinfo {author} {\bibfnamefont {C.}~\bibnamefont
  {Filgueiras}}\ and\ \bibinfo {author} {\bibfnamefont {F.}~\bibnamefont
  {Moraes}},\ }\href {\doibase 10.1016/j.aop.2008.08.002} {\bibfield  {journal}
  {\bibinfo  {journal} {Ann. Phys. (N.Y.)}\ }\textbf {\bibinfo {volume}
  {323}},\ \bibinfo {pages} {3150} (\bibinfo {year} {2008})}\BibitemShut
  {NoStop}%
\bibitem [{\citenamefont {Katanaev}\ and\ \citenamefont
  {Volovich}(1992)}]{AP.1992.216.1}%
  \BibitemOpen
  \bibfield  {author} {\bibinfo {author} {\bibfnamefont {M.}~\bibnamefont
  {Katanaev}}\ and\ \bibinfo {author} {\bibfnamefont {I.}~\bibnamefont
  {Volovich}},\ }\href {\doibase 10.1016/0003-4916(52)90040-7} {\bibfield
  {journal} {\bibinfo  {journal} {Ann. Phys. (N.Y.)}\ }\textbf {\bibinfo
  {volume} {216}},\ \bibinfo {pages} {1} (\bibinfo {year} {1992})}\BibitemShut
  {NoStop}%
\bibitem [{\citenamefont {Azevedo}\ and\ \citenamefont
  {Moraes}(2000)}]{JP.2000.12.7421}%
  \BibitemOpen
  \bibfield  {author} {\bibinfo {author} {\bibfnamefont {S.}~\bibnamefont
  {Azevedo}}\ and\ \bibinfo {author} {\bibfnamefont {F.}~\bibnamefont
  {Moraes}},\ }\href {\doibase 1088/0953-8984/12/33/309} {\bibfield  {journal}
  {\bibinfo  {journal} {J. Phys.: Condens. Matter}\ }\textbf {\bibinfo {volume}
  {12}},\ \bibinfo {pages} {7421} (\bibinfo {year} {2000})}\BibitemShut
  {NoStop}%
\bibitem [{\citenamefont {Miranda}\ and\ \citenamefont
  {Moraes}(2003)}]{JPA.2003.36.863}%
  \BibitemOpen
  \bibfield  {author} {\bibinfo {author} {\bibfnamefont {J.~A.}\ \bibnamefont
  {Miranda}}\ and\ \bibinfo {author} {\bibfnamefont {F.}~\bibnamefont
  {Moraes}},\ }\href {\doibase 10.1088/0305-4470/36/3/319} {\bibfield
  {journal} {\bibinfo  {journal} {J. Phys. A}\ }\textbf {\bibinfo {volume}
  {36}},\ \bibinfo {pages} {863} (\bibinfo {year} {2003})}\BibitemShut
  {NoStop}%
\bibitem [{\citenamefont {Osipov}\ and\ \citenamefont
  {Kochetov}(2001)}]{JETP.2001.73.562}%
  \BibitemOpen
  \bibfield  {author} {\bibinfo {author} {\bibfnamefont {V.}~\bibnamefont
  {Osipov}}\ and\ \bibinfo {author} {\bibfnamefont {E.}~\bibnamefont
  {Kochetov}},\ }\href {\doibase 10.1134/1.1387528} {\bibfield  {journal}
  {\bibinfo  {journal} {JETP Lett.}\ }\textbf {\bibinfo {volume} {73}},\
  \bibinfo {pages} {562} (\bibinfo {year} {2001})},\ \bibinfo {note}
  {10.1134/1.1387528}\BibitemShut {NoStop}%
\bibitem [{\citenamefont {de~M.~Carvalho}\ \emph {et~al.}(2007)\citenamefont
  {de~M.~Carvalho}, \citenamefont {S\'{a}tiro},\ and\ \citenamefont
  {Moraes}}]{EPL.2007.80.46002}%
  \BibitemOpen
  \bibfield  {author} {\bibinfo {author} {\bibfnamefont {A.~M.}\ \bibnamefont
  {de~M.~Carvalho}}, \bibinfo {author} {\bibfnamefont {C.}~\bibnamefont
  {S\'{a}tiro}}, \ and\ \bibinfo {author} {\bibfnamefont {F.}~\bibnamefont
  {Moraes}},\ }\href {\doibase 10.1209/0295-5075/80/46002} {\bibfield
  {journal} {\bibinfo  {journal} {Europhys. Lett.}\ }\textbf {\bibinfo {volume}
  {80}},\ \bibinfo {pages} {46002} (\bibinfo {year} {2007})}\BibitemShut
  {NoStop}%
\bibitem [{\citenamefont {Kowalski}\ \emph {et~al.}(2002)\citenamefont
  {Kowalski}, \citenamefont {Podlaski},\ and\ \citenamefont
  {Rembieli\'{n}ski}}]{PRA.2002.66.032118}%
  \BibitemOpen
  \bibfield  {author} {\bibinfo {author} {\bibfnamefont {K.}~\bibnamefont
  {Kowalski}}, \bibinfo {author} {\bibfnamefont {K.}~\bibnamefont {Podlaski}},
  \ and\ \bibinfo {author} {\bibfnamefont {J.}~\bibnamefont
  {Rembieli\'{n}ski}},\ }\href {\doibase 10.1103/PhysRevA.66.032118} {\bibfield
   {journal} {\bibinfo  {journal} {Phys. Rev. A}\ }\textbf {\bibinfo {volume}
  {66}},\ \bibinfo {pages} {032118} (\bibinfo {year} {2002})}\BibitemShut
  {NoStop}%
\bibitem [{\citenamefont {Reed}\ and\ \citenamefont
  {Simon}(1975)}]{Book.1975.Reed.II}%
  \BibitemOpen
  \bibfield  {author} {\bibinfo {author} {\bibfnamefont {M.}~\bibnamefont
  {Reed}}\ and\ \bibinfo {author} {\bibfnamefont {B.}~\bibnamefont {Simon}},\
  }\href@noop {} {\emph {\bibinfo {title} {Methods of Modern Mathematical
  Physics. II. Fourier Analysis, Self-Adjointness.}}}\ (\bibinfo  {publisher}
  {Academic Press},\ \bibinfo {address} {New York - London},\ \bibinfo {year}
  {1975})\BibitemShut {NoStop}%
\bibitem [{\citenamefont {Kay}\ and\ \citenamefont
  {Studer}(1991)}]{CMP.1991.139.103}%
  \BibitemOpen
  \bibfield  {author} {\bibinfo {author} {\bibfnamefont {B.~S.}\ \bibnamefont
  {Kay}}\ and\ \bibinfo {author} {\bibfnamefont {U.~M.}\ \bibnamefont
  {Studer}},\ }\href {\doibase 10.1007/BF02102731} {\bibfield  {journal}
  {\bibinfo  {journal} {Commun. Math. Phys.}\ }\textbf {\bibinfo {volume}
  {139}},\ \bibinfo {pages} {103} (\bibinfo {year} {1991})}\BibitemShut
  {NoStop}%
\bibitem [{\citenamefont {Hayashi}(2005)}]{PLA.2005.342.237}%
  \BibitemOpen
  \bibfield  {author} {\bibinfo {author} {\bibfnamefont {M.}~\bibnamefont
  {Hayashi}},\ }\href {\doibase 10.1016/j.physleta.2005.05.037} {\bibfield
  {journal} {\bibinfo  {journal} {Phys. Lett. A}\ }\textbf {\bibinfo {volume}
  {342}},\ \bibinfo {pages} {237} (\bibinfo {year} {2005})}\BibitemShut
  {NoStop}%
\bibitem [{\citenamefont {Bonneau}\ \emph {et~al.}(2001)\citenamefont
  {Bonneau}, \citenamefont {Faraut},\ and\ \citenamefont
  {Valent}}]{AJP.2001.69.322}%
  \BibitemOpen
  \bibfield  {author} {\bibinfo {author} {\bibfnamefont {G.}~\bibnamefont
  {Bonneau}}, \bibinfo {author} {\bibfnamefont {J.}~\bibnamefont {Faraut}}, \
  and\ \bibinfo {author} {\bibfnamefont {G.}~\bibnamefont {Valent}},\ }\href
  {\doibase 10.1119/1.1328351} {\bibfield  {journal} {\bibinfo  {journal} {Am.
  J. Phys.}\ }\textbf {\bibinfo {volume} {69}},\ \bibinfo {pages} {322}
  (\bibinfo {year} {2001})}\BibitemShut {NoStop}%
\bibitem [{not()}]{note-nrd}%
  \BibitemOpen
  \href@noop {} {}\bibinfo {note} {These results are different from those in
  reference \cite{AP.2008.323.3150} since in equation (37) in this reference
  has a sign mistake.}\BibitemShut {Stop}%
\bibitem [{\citenamefont {Andrade}\ \emph {et~al.}(2012)\citenamefont
  {Andrade}, \citenamefont {Silva},\ and\ \citenamefont
  {Pereira}}]{PRD.2012.85.041701}%
  \BibitemOpen
  \bibfield  {author} {\bibinfo {author} {\bibfnamefont {F.~M.}\ \bibnamefont
  {Andrade}}, \bibinfo {author} {\bibfnamefont {E.~O.}\ \bibnamefont {Silva}},
  \ and\ \bibinfo {author} {\bibfnamefont {M.}~\bibnamefont {Pereira}},\ }\href
  {\doibase 10.1103/PhysRevD.85.041701} {\bibfield  {journal} {\bibinfo
  {journal} {Phys. Rev. D}\ }\textbf {\bibinfo {volume} {85}},\ \bibinfo
  {pages} {041701(R)} (\bibinfo {year} {2012})}\BibitemShut {NoStop}%
\bibitem [{\citenamefont {Filgueiras}\ \emph {et~al.}(2010)\citenamefont
  {Filgueiras}, \citenamefont {Silva}, \citenamefont {Oliveira},\ and\
  \citenamefont {Moraes}}]{AP.2010.325.2529}%
  \BibitemOpen
  \bibfield  {author} {\bibinfo {author} {\bibfnamefont {C.}~\bibnamefont
  {Filgueiras}}, \bibinfo {author} {\bibfnamefont {E.~O.}\ \bibnamefont
  {Silva}}, \bibinfo {author} {\bibfnamefont {W.}~\bibnamefont {Oliveira}}, \
  and\ \bibinfo {author} {\bibfnamefont {F.}~\bibnamefont {Moraes}},\ }\href
  {\doibase 10.1016/j.aop.2010.05.012} {\bibfield  {journal} {\bibinfo
  {journal} {Ann. Phys. (N.Y.)}\ }\textbf {\bibinfo {volume} {325}},\ \bibinfo
  {pages} {2529} (\bibinfo {year} {2010})}\BibitemShut {NoStop}%
\bibitem [{\citenamefont {Furtado}\ \emph {et~al.}(2007)\citenamefont
  {Furtado}, \citenamefont {Rosas},\ and\ \citenamefont
  {Azevedo}}]{EPL.2007.79.57001}%
  \BibitemOpen
  \bibfield  {author} {\bibinfo {author} {\bibfnamefont {C.}~\bibnamefont
  {Furtado}}, \bibinfo {author} {\bibfnamefont {A.}~\bibnamefont {Rosas}}, \
  and\ \bibinfo {author} {\bibfnamefont {S.}~\bibnamefont {Azevedo}},\ }\href
  {\doibase 10.1209/0295-5075/79/57001} {\bibfield  {journal} {\bibinfo
  {journal} {Europhys. Lett.}\ }\textbf {\bibinfo {volume} {79}},\ \bibinfo
  {pages} {57001} (\bibinfo {year} {2007})}\BibitemShut {NoStop}%
\end{thebibliography}
\end{document}